\documentclass[11pt,onecolumn]{article}
\usepackage{geometry,setspace}
\usepackage[font=small,format=plain,labelfont=bf,justification=raggedright,singlelinecheck=false]{caption}
\usepackage{hyperref,cite}
\usepackage{amsmath,amsfonts,amssymb,amsthm,upgreek,mathrsfs,bm}
\usepackage{graphicx,subfig,tikz}
\usepackage{physics}
\usepackage{extarrows}
\usepackage{orcidlink}

\hypersetup{colorlinks=true,linkcolor=blue,anchorcolor=blue,citecolor=blue,urlcolor=blue}
\allowdisplaybreaks[2]

\numberwithin{equation}{section}

\makeatletter
\newcommand{\directint}{\mathop{\mathpalette\direct@int\relax}\!\int}
\newcommand{\direct@int}[2]{%
	\sbox\z@{%
		\m@th
		\ooalign{%
			\hidewidth$\demote@style{#1}{\bigoplus}$\hidewidth\cr
			$#1\phantom{\int}$\cr
		}%
	}%
	\wd\z@=\z@\box\z@
}
\newcommand{\demote@style}[2]{%
	\ifx#1\displaystyle\scriptstyle#2\else
		\raise.5\fontdimen22\textfont2\hbox{$\scriptscriptstyle#2$}%
	\fi
}
\makeatother

\title{Carrollian Dictionary for Massive Particles at Null Infinity}
\author{Yu-fan Zheng\orcidlink{0000-0001-7405-582X}$^{a}$\footnote{\href{mailto:zhengyufan@bimsa.cn}{zhengyufan@bimsa.cn}}}
\date{\today}

\begin{document}

	\maketitle
	\begin{center}
		{\itshape
			$^{a}$ Beijing Institute of Mathematical Sciences and Applications (BIMSA), \\
			Huaibei Town, Huairou District, Beijing 101408, China \\
		}
		\vspace{10mm}
	\end{center}

	\begin{abstract}
		Massive particles reach timelike rather than null infinity, leaving their Carrollian boundary description unresolved.
		We construct a Carrollian dictionary for massive one-particle states on $\mathscr I^\pm$ without requiring their worldlines to reach null infinity.
		The dictionary encodes the bulk momentum through its projections onto the null frame associated with each celestial point.
		The Poincar\'e intertwining equations require the complete Carrollian representation to describe massive states.
		We apply the dictionary to the K\"all\'en--Lehmann representation and to soft photon and graviton theorems.
		The K\"all\'en--Lehmann application fixes the coefficient function of the non-contact two-point function in terms of the bulk spectral density.
		Soft theorems show that global modes reproduce the ordinary $U(1)$ and Poincar\'e actions, whereas local actions remain integrals over celestial directions. \par
	\end{abstract}

	\tableofcontents

\section{Introduction}

	Null infinity is the natural arena for massless radiation, BMS symmetry, and soft theorems \cite{Bondi:1962px,Sachs:1962wk,Strominger:2013jfa,Strominger:2017zoo}.
	Its radiative phase space provides the asymptotic boundary data for massless scattering \cite{Ashtekar:1981bq}.
	Celestial and Carrollian formulations express this structure through boundary states on $\mathscr I^\pm$ and Ward identities for their hard and soft sectors \cite{Pasterski:2016qvg,Pasterski:2017kqt,Donnay:2022aba,Donnay:2022wvx,Nguyen:2023vfz,Nguyen:2023miw,Saha:2023hsl,Ruzziconi:2026bix,Zhu:2026ofh,Bagchi:2022emh,Liu:2024nkc,Bekaert:2024itn}.
	Massive particles, however, reach timelike rather than null infinity.
	Although a Carrollian description of massive asymptotic states has been constructed at the blow-up of timelike infinity \cite{Have:2024dff,Borthwick:2024skd,Liu:2025oom}, a counterpart at null infinity remains unknown.
	Several recent works have explored related approaches to bulk reconstruction and to describing massive particles from asymptotic boundaries \cite{Chen:2023naw,Hao:2025btl,Hao:2026cqm,Melton:2026tdw}.
	Accordingly, what is missing is a normalized map that intertwines the bulk and boundary Poincar\'e actions, encodes massive one-particle states in a Carrollian state space at null infinity, and allows the bulk state to be reconstructed. \par

	In this letter, we solve the Poincar\'e intertwining equations to construct the map in eq. \eqref{eq:Dictionary}, which relates a 4D bulk to its 3D null boundary.
	The Carrollian representation commonly used for massless radiation has vanishing quadratic Poincar\'e Casimir and therefore cannot describe massive particles.
	By contrast, the complete representation introduced in \cite{Zheng:2026kuf} contains representation orbits with nonzero quadratic Casimir, and the intertwining equations admit real solutions for the representation labels of generic massive momenta.
	In the resulting dictionary, the celestial coordinates $z^a$ specify the null frame in eq. \eqref{eq:NullFrame} onto which the massive momentum is projected.
	The delta function support identifies $\beta^a$ and $\kappa$ as the corresponding projection coefficients.
	In this dictionary, the momentum information is distributed over the whole celestial sphere rather than being localized at a single boundary point, as in the massless case. \par

	The dictionary is an isometric embedding into an overcomplete boundary state space, while applying the reconstruction map followed by the dictionary yields a nontrivial projection operator on that space.
	The construction extends to arbitrary spin through a unitary spin matrix factor, whose general form is given in Appendix \ref{app:GeneralSpin}.
	Applying the map to the K\"all\'en--Lehmann representation fixes the coefficient function of the generic non-contact Carrollian two-point function in terms of the bulk spectral density.
	The dictionary also determines how soft theorems act on massive hard states.
	Unlike the massless case, where an angular pole evaluates the smearing function at the hard leg's celestial direction, the smearing is not localized for a massive hard leg.
	The actions of local large gauge transformations and supertranslations remain Poisson transforms over the celestial sphere, and the projection picture indicates an analogous structure for local superrotations.
	By contrast, the global modes combine the angular dependence and reproduce the ordinary $U(1)$ and Poincar\'e actions.
	Together with the standard radiative dictionary, this provides a common Carrollian framework for massive and massless external states on $\mathscr I^\pm$. \par

\section{Massive Carrollian Dictionary and Reconstruction}\label{sec:MassiveDictionary}

	We construct the dictionary and reconstruction maps between the massive one-particle Hilbert space and the state space of the complete Carrollian representation.
	The dictionary gives an isometric embedding into an overcomplete boundary state space.
	In the bulk, we use the Minkowski metric $\eta_{\mu\nu}=\mathrm{diag}(-,+,+,+)$ and the asymptotic momentum eigenstates $\ket{p;j,s}$.
	At fixed spin $j$, we define the outgoing and incoming Hilbert spaces by the direct integrals over the mass spectrum
	\begin{equation}
		\mathcal{H}_{j}^{(\pm)}=\directint_{\mu^2>0}\dd{\mu^2}\,\mathcal{H}_{\mu,j}^{(\pm)}.
	\end{equation}
	The superscripts $(+)$ and $(-)$ denote outgoing and incoming particles, respectively. \par

	On the boundary, we use the state space of the complete Carrollian representation \cite{Chen:2020vvn,Chen:2021xkw,Chen:2022cpx,Chen:2023pqf,Zheng:2026kuf}.
	A boundary ket state is $\ket{O_X}=\ket{u,z^a;\Delta,l,\beta^a,\kappa}$, with $X=(u,z^a;\Delta,l,\beta^a,\kappa)$ and $a,b=1,2$.
	Here $(u,z^a)$ are Carrollian coordinates on null infinity, whereas $(\Delta,l,\beta^a,\kappa)$ are representation labels.
	The complete representation contains one orbit with $\kappa>0$, one with $\kappa<0$, and $\kappa=0$ orbits labeled by fixed $\vec{\beta}^{\,2}$.
	The Carrollian representation commonly used for massless radiation lies in the $\vec{\beta}=\kappa=0$ sector and has $\mathcal{C}_2=0$, making it suitable for the standard dictionary for massless particles.
	The first two orbits have the nonzero quadratic Casimir required for massive states, and we denote the state spaces that they span by
	\begin{equation}
		\mathcal{S}_O^{>0}=\mathrm{span}\left\{\ket{O_X}\mid\kappa>0\right\},
		\qquad
		\mathcal{S}_O^{<0}=\mathrm{span}\left\{\ket{O_X}\mid\kappa<0\right\}.
	\end{equation}
	In what follows, we present the outgoing map explicitly.
	The incoming map follows from the analogous construction relating $\mathcal{H}_{j}^{(-)}$ to $\mathcal{S}_O^{<0}$. \par

	To connect the two representations, we introduce the following null frame on the celestial sphere
	\begin{equation}\label{eq:NullFrame}
		\begin{aligned}
			q_0^\mu(z)&=(1+\vec{z}^{\,2},2z^1,2z^2,1-\vec{z}^{\,2}),\\
			q_1^\mu(z)&=\frac{1}{2}\partial_1 q_0^\mu(z) = (z^1,1,0,-z^1),\\
			q_2^\mu(z)&=\frac{1}{2}\partial_2 q_0^\mu(z) = (z^2,0,1,-z^2),\\
			q_3^\mu(z)&=\frac{1}{8}\partial^2 q_0^\mu(z) = \frac{1}{2}(1,0,0,-1),
		\end{aligned}
	\end{equation}
	where $z^a$ are stereographic coordinates and $\vec{z}^{\,2}=z^a z^a$.
	The relations among the frame vectors are collected in Appendix \ref{app:BulkBoundaryParameters}. \par

	Matching the action of the Poincar\'e generators on the bulk and boundary states fixes the scalar dictionary kernel $\mathcal{G}(p;X)$ up to normalization.
	For a scalar bulk state, the dictionary acts as
	\begin{equation}\label{eq:Dictionary}
		\ket*{O_X^{l=0}}=\hat{\mathcal{G}}\ket{p}=\int_{\substack{p^2<0\\p^0>0}}\frac{\dd[4]{p}}{(2\pi)^2}\,\ket{p}\mathcal{G}(p;X).
	\end{equation}
	The dictionary kernel is given by
	\begin{equation}
		\begin{aligned}
			&\mathcal{G}(p;X)
			={}2m\,
			e^{2iu\,p\cdot q_3(z)}
			\abs{\kappa}^{1-\Delta}
			\times\delta^{(2)}\!\left(\beta^a+p\cdot q_a(z)\right)
			\delta\!\left(\kappa+p\cdot q_0(z)\right).
		\end{aligned}
	\end{equation}
	Here $m^2=-p^2$, with $m$ denoting the mass of the bulk particle.
	The overall factor $2m$ is fixed by the normalization of the dictionary map.
	The delta function constraints in the scalar dictionary kernel require
	\begin{equation}
		\beta^a=-p\cdot q_a(z),
		\qquad
		\kappa=-p\cdot q_0(z).
	\end{equation}
	Thus $\beta^a$ and $\kappa$ acquire their physical interpretation as the coefficients obtained by projecting the bulk momentum onto the null frame.
	The momentum is reconstructed entirely from $m$, $z^a$, $\beta^a$, and $\kappa$ as\footnote{Here $\rho\equiv(m^2+\vec{\beta}^{\,2})/\kappa=p^0+p^3=-2p\cdot q_3$ is Fourier conjugate to $u$, since the kernel phase is $e^{2iu p\cdot q_3}=e^{-iu\rho}$.
		In terms of $\rho$, the momentum and quadratic Casimir take the forms $p^\mu=\frac{\rho}{2}q_0^\mu(z)-\beta^a q_a^\mu(z)+\kappa q_3^\mu(z)$ and $\mathcal{C}_2=\kappa\rho-\vec{\beta}^{\,2}=m^2$.}
	\begin{equation}\label{eq:MomentumFrame}
		p^\mu=\frac{m^2+\vec{\beta}^{\,2}}{2\kappa}q_0^\mu(z)-\beta^a q_a^\mu(z)+\kappa q_3^\mu(z).
	\end{equation}
	Outgoing momenta directed toward the future have $\kappa>0$.
	When all external momenta are taken as outgoing, incoming particles are represented by momenta directed toward the past and have $\kappa<0$.
	The corresponding states therefore lie in $\mathcal{S}_O^{>0}$ and $\mathcal{S}_O^{<0}$, respectively. \par

	The scalar reconstruction map $\hat{\mathcal{K}}:\mathcal{S}_O^{>0}\to\mathcal{H}_{j}^{(+)}$ is given by
	\begin{equation}
		\ket{p}=\hat{\mathcal{K}}\ket{O_X}=\int_{\kappa>0}\dd{u}\dd[2]{z}\dd[2]{\beta}\dd{\kappa}\,\ket*{O_X^{l=0}}\mathcal{K}(X;p).
	\end{equation}
	The reconstruction kernel is given by
	\begin{equation}
		\begin{aligned}
			&\mathcal{K}(X;p)
			={}2m\,
			e^{-2iu\,p\cdot q_3(z)}
			\abs{\kappa}^{\Delta-3}
			\times
			\delta^{(2)}\!\left(\beta^a+p\cdot q_a(z)\right)
			\delta\!\left(\kappa+p\cdot q_0(z)\right).
		\end{aligned}
	\end{equation}
	\par

	On the unitary line $\Delta=2+i\nu$ with $\nu\in\mathbb{R}$,\footnote{Because $\kappa$ carries conformal dimension and the pairing includes $\dd{\kappa}$, this unitary line is shifted relative to the principal series used in celestial CFT \cite{Iacobacci:2024laa,Liu:2021tif}.} the scalar kernels obey
	\begin{equation}
		\mathcal{K}(X;p)=\mathcal{G}(p;X)^*.
	\end{equation}
	For spin $j$, the scalar support is unchanged, while the kernels take the forms $\mathcal{G}_{ls}^{(j)}(p;X)=\mathcal{G}(p;X)\mathcal{R}_{ls}^{(j)}$ and $\mathcal{K}_{sl}^{(j)}(X;p)=\mathcal{K}(X;p)(\mathcal{R}^{(j)\dagger})_{sl}$, where $s$ and $l$ are the bulk and boundary spin labels.
	The spin matrix factor, the corresponding kernels for arbitrary spin, and the relation $\mathcal{K}^{(j)}=\mathcal{G}^{(j)\dagger}$ are given in Appendix \ref{app:GeneralSpin}.
	Therefore, the reconstruction map is the Hermitian adjoint of the dictionary map between $\mathcal{H}_{j}^{(+)}$ and $\mathcal{S}_O^{>0}$ and obeys
	\begin{equation}
		\hat{\mathcal{K}}=\hat{\mathcal{G}}^\dagger.
	\end{equation}
	\par

	Applying $\bra{p}$ to the ket dictionary in eq. \eqref{eq:Dictionary} and its bra counterpart to $\ket{p}$, with $\bra{p^\prime}\ket{p}=(2\pi)^2\delta^{(4)}(p-p^\prime)$, gives
	\begin{equation}
		\mathcal{G}(p;X)=\bra{p}\ket{O_X},
		\qquad
		\mathcal{K}(X;p)=\bra{O_X}\ket{p}.
	\end{equation}
	These kernels are state overlaps rather than bulk-to-boundary propagators, because massive worldlines do not reach null infinity.
	The dictionary and reconstruction nevertheless have an integral structure analogous to an HKLL transform at the level of states. \par

	Applying $\bra{p}$ to the reconstruction of $\ket{p^\prime}$ and using the dictionary overlap gives
	\begin{equation}
		(2\pi)^2\delta^{(4)}(p-p^\prime)=\int_{\kappa>0}\dd{u}\dd[2]{z}\dd[2]{\beta}\dd{\kappa}\,\mathcal{K}(X;p^\prime)\mathcal{G}(p;X).
	\end{equation}
	The two compositions therefore obey
	\begin{equation}
		\begin{aligned}
			&\hat{\mathcal{G}}^\dagger\hat{\mathcal{G}}=\mathbf{I}_{\mathcal{H}_{j}^{(+)}}, \qquad
			\hat{\mathcal{G}}\hat{\mathcal{G}}^\dagger=\Pi_{\mathcal{S}_O^{>0}}\neq\mathbf{I}_{\mathcal{S}_O^{>0}}.
		\end{aligned}
	\end{equation}
	The first identity makes the dictionary an isometric embedding, while the second shows that the reverse composition is a nontrivial projection operator on $\mathcal{S}_O^{>0}$.
	The null frame and $\Delta$ provide redundant labels, so $\ker\hat{\mathcal{G}}^\dagger\neq0$ and the reverse composition is not the identity.
	More precisely, the second composition is a projection operator because
	\begin{equation}
		\Pi_{\mathcal{S}_O^{>0}}^2=\hat{\mathcal{G}}(\hat{\mathcal{G}}^\dagger\hat{\mathcal{G}})\hat{\mathcal{G}}^\dagger=\hat{\mathcal{G}}\hat{\mathcal{G}}^\dagger=\Pi_{\mathcal{S}_O^{>0}}.
	\end{equation}
	Through the dictionary, the operators $O_X$ form an overcomplete basis for $\mathcal{H}_{j}^{(+)}$.
	Inserting the resolution of the identity in $\mathcal{S}_O^{>0}$ rewrites the isometry condition on $\mathcal{H}_{j}^{(+)}$ as
	\begin{equation}
		\mathbf{I}_{\mathcal{H}_{j}^{(+)}}=\int_{\kappa>0}\dd{u}\dd[2]{z}\dd[2]{\beta}\dd{\kappa}\,\hat{\mathcal{G}}^\dagger\ket{O_X}\bra{O_X}\hat{\mathcal{G}}.
	\end{equation}
	Thus the pullbacks of the Carrollian states form a Parseval continuous frame for $\mathcal{H}_{j}^{(+)}$. \par

\section{Spectral Data in the Two-Point Function}

	One physical consequence of the massive dictionary is that it identifies the dynamical data left undetermined by Carrollian symmetry.
	For a scalar bulk field, the momentum space K\"all\'en--Lehmann representation of the Wightman two-point function takes the form \cite{Kallen:1952zz,Lehmann:1954xi,Sleight:2023ojm,Iacobacci:2024nhw,Pacifico:2024dyo}
	\begin{equation}
		\begin{aligned}
			&\ev{\Phi(p_1)\Phi(p_2)}
			={}(2\pi)^2\delta^{(4)}(p_1+p_2)
			\times\int_0^\infty \dd{\mu^2}
			\rho_{\mathrm{KL}}(\mu^2)\,
			\theta(p_1^0)\delta(p_1^2+\mu^2),
		\end{aligned}
	\end{equation}
	where $\rho_{\mathrm{KL}}(\mu^2)$ is the bulk spectral density. \par

	Applying the massive dictionary to both bulk legs gives the Carrollian correlator.
	On the massive mass shell, the positive energy condition imposed by $\theta(p_1^0)$ is equivalently expressed as $\theta(\kappa_1)$.
	Since the first insertion belongs to the $\kappa_1>0$ orbit, this factor is unity and is omitted below.
	The resulting Carrollian correlator is
	\begin{equation}
		\begin{aligned}
			&\ev{O_{X_1}O_{X_2}}_{\mathrm{KL}}
			=\frac{\mu^2}{2\pi^2}\rho_{\mathrm{KL}}(\mu^2)\\
			&\qquad\times e^{-i(u_1-u_2)\frac{(\vec{\beta}_1^{\,2}-\vec{\beta}_2^{\,2})}{\kappa_1+\kappa_2}}
			\frac{|\kappa_1|^{1-\Delta_1}|\kappa_2|^{1-\Delta_2}}{|\vec{\beta}_1^{\,2}-\vec{\beta}_2^{\,2}|}\\
			&\qquad\times
			\delta^{(2)}\left((z^a_1-z^a_2)-\frac{(\kappa_1+\kappa_2)(\beta^a_1+\beta^a_2)}{\vec{\beta}_1^{\,2}-\vec{\beta}_2^{\,2}}\right).
		\end{aligned}
	\end{equation}
	Here the delta function support fixes the bulk mass variable as
	\begin{equation}
		\mu^2=-\frac{\kappa_2\vec{\beta}_1^{\,2}+\kappa_1\vec{\beta}_2^{\,2}}{\kappa_1+\kappa_2}.
	\end{equation}
	For a massive channel, $\mu^2>0$ together with $\kappa_1>0$ requires $\kappa_2<0$, so the first and second insertions belong to the outgoing and incoming orbits identified in Section \ref{sec:MassiveDictionary}, respectively. \par

	The resulting correlator has the same kinematic form and distributional support as the generic scalar non-contact two-point function obtained from the global Ward identities in \cite{Zheng:2026kuf}.
	The general scalar two-point function obtained from the Carrollian Ward identities contains an undetermined coefficient function $f(\mu^2,L)$.
	Matching this solution to the correlator above fixes it to
	\begin{equation}
		f(\mu^2,L)=\frac{\mu^2}{2\pi^2}\rho_{\mathrm{KL}}(\mu^2).
	\end{equation}
	Here $L=(\vec{z}_1-\vec{z}_2)^2/(\kappa_1\kappa_2)$ is the second invariant.
	The bulk spectral density for scalars therefore fixes $f$ and introduces no independent dependence on $L$.
	For a unitary bulk theory, spectral positivity further requires $f(\mu^2,L)\geq0$ for $\mu^2>0$. \par

\section{Soft Theorems with Massive External Legs}

	We use soft theorems for amplitudes containing one outgoing soft photon or graviton and $n$ massive scalar hard legs to determine how the corresponding boundary symmetries act on massive states.
	The interpretation of soft theorems as Ward identities of asymptotic symmetries has been developed in several complementary formulations \cite{Strominger:2013jfa,Agrawal:2025bsy,Isen:2026xoc,Campiglia:2015lxa,Melton:2026tdw}.
	Within this correspondence, the soft photon theorem determines the action of large gauge transformations \cite{Weinberg:1965nx,Campiglia:2015qka,Strominger:2017zoo}, while the leading and subleading soft graviton theorems reproduce translations and Lorentz rotations, respectively \cite{He:2014laa,Campiglia:2015kxa,Cachazo:2014fwa,Campiglia:2014yka}.
	We test these identifications below using the massive dictionary defined in Section \ref{sec:MassiveDictionary}. \par

	At a given soft order, write $J_h$ for the corresponding current and $K^h(w)$ for its smearing kernel.
	The associated soft charge is
	\begin{equation}\label{eq:SoftCharge}
		Q_S=\sum_h\int \dd[2]{w}\,K^h(w)J_h.
	\end{equation}
	The soft theorem also determines the hard action $\mathcal{I}_{A}$, namely the action of the hard charge on leg $A$.
	The corresponding Ward identity takes the general form
	\begin{equation}\label{eq:SoftWardIdentity}
		\ev{Q_S\,{\textstyle\prod}_{A=1}^{n}O_A}=\sum_{A=1}^{n}\ev{O_1\cdots(\mathcal{I}_{A}O_A)\cdots O_n}.
	\end{equation}
	The soft charge inserts the corresponding radiative mode, while $\mathcal{I}_{A}$ gives the hard actions.
	The Ward identity equates this soft insertion with the sum of the hard actions.
	Here $\eta_A=\pm1$ for outgoing and incoming legs, respectively.
	We use the anti-Hermitian differential realization of the Carrollian generators and write the hard action as $\mathcal{I}_{A}O_A=i\eta_A\delta O_A$. \par

	For an outgoing helicity-$h$ soft photon, the modified Mellin transform acts only on its energy $\omega$ and is given by \cite{Donnay:2022wvx,Nguyen:2023miw}
	\begin{equation}
		\mathcal{O}_{\Delta,h}(u,w)=
		\int_0^\infty \dd{\omega}\,\omega^{\Delta-1}e^{-2iu\omega}a_h^{\mathrm{out}}(\omega,w).
	\end{equation}
	The soft momentum is $p_s^\mu=\omega q_0^\mu(w)$.
	We use the polarization vector $\varepsilon_h^\mu(w)=q_1^\mu(w)+i\,\operatorname{sgn}(h)q_2^\mu(w)$ for both photons and gravitons.
	In the Carrollian dictionary, this leg lies on the orbit $\beta^a=\kappa=0$, while $z^a=w^a$ labels its celestial direction.
	For the leading soft photon, the pole at $\Delta=1$ defines the Carrollian current \cite{Donnay:2018neh}
	\begin{equation}
		J_h(u,w)=\lim_{\Delta\to1}(\Delta-1)\mathcal{O}_{\Delta,h}(u,w).
	\end{equation}
	The leading soft photon theorem associates the factor $S_{\gamma,A}^{h}(w)=e_A(p_A\cdot\varepsilon_h(w))/(p_A\cdot q_0(w))$ with hard leg $A$, where $e_A$ is its electric charge.
	Substituting the massive momentum parametrization in eq. \eqref{eq:MomentumFrame} gives the contribution of the hard leg,
	\begin{equation}
		S_{\gamma,A}^{h}(w)=e_A
		\frac{\kappa_A(\beta_A^1+ih\beta_A^2)+(m_A^2+\vec{\beta}_A^{\,2})((w^1-z_A^1)+ih(w^2-z_A^2))}
		{\kappa_A^2+2\kappa_A\beta_A^a(w^a-z_A^a)+(m_A^2+\vec{\beta}_A^{\,2})(\vec{w}-\vec{z}_A)^2}.
	\end{equation}
	Unlike the massless case, this factor is not an angular pole, so its smearing does not reduce to evaluating the gauge parameter at a single celestial direction. \par

	For a large gauge parameter $\alpha(w)$, the photon smearing kernel is
	\begin{equation}
		K_\alpha^h(w)=-\frac{1}{4\pi}(\partial_1-ih\partial_2)\alpha(w).
	\end{equation}
	Substituting $K^h=K_\alpha^h$ into eq. \eqref{eq:SoftCharge} gives the soft photon charge $Q_S^\gamma[\alpha]$.
	Assuming that the boundary term in stereographic coordinates vanishes, integration by parts gives the massive hard action as the Poisson transform
	\begin{equation}
		\mathcal{I}_{A}^{\gamma}[\alpha]=\eta_A\frac{e_A m_A^2}{\pi}\int \dd[2]{w}\,\frac{\alpha(w)}{(-p_A\cdot q_0(w))^2}.
	\end{equation}
	For a constant parameter $\alpha_0$, the photon smearing kernel vanishes, so this global mode is not fixed by the integration by parts relation.
	Global $U(1)$ symmetry instead fixes the multiplicative hard action $\mathcal{I}_{A}^{\gamma}[\alpha_0]O_A=\eta_Ae_A\alpha_0O_A$, and the Ward identity reduces to charge conservation.
	A local $\alpha(w)$ acts through the nonlocal Poisson transform above. \par

	For an outgoing leading soft graviton, the same $\Delta=1$ residue with $h=\pm2$ defines $J_h$.
	Omitting the conventional overall factor $\sqrt{32\pi G}/2$ throughout the rest of this letter, the leading soft factor for hard leg $A$ is $S_{g,A}^{(0),h}(w)=(p_A\cdot\varepsilon_h(w))^2/(p_A\cdot q_0(w))$.
	For a supertranslation parameter $f(w)$, the graviton smearing kernel is
	\begin{equation}
		K_f^h(w)=-\frac{1}{8\pi}(\partial_1-i\,\operatorname{sgn}(h)\partial_2)^2f(w).
	\end{equation}
	Substituting $K^h=K_f^h$ into eq. \eqref{eq:SoftCharge} gives the supertranslation soft charge $Q_S^g[f]$.
	The corresponding massive hard action entering eq. \eqref{eq:SoftWardIdentity} is
	\begin{equation}
		\mathcal{I}_{A}^{g}[f]=\eta_A\frac{2m_A^4}{\pi}\int \dd[2]{w}\,\frac{f(w)}{(-p_A\cdot q_0(w))^3}.
	\end{equation}
	For $f\in\operatorname{span}\{1,w^a,-\vec{w}^{\,2}\}$, the graviton smearing kernel vanishes, so these global modes are not fixed by the integration by parts relation.
	The ordinary translation representation instead fixes the hard action on these four global modes, and the Ward identity reduces to momentum conservation.
	On this global subspace, $\mathcal{I}_{A}^{g}[f]O_X=i\eta_A\delta_fO_X$, where
	\begin{equation}
		\delta_f O_X=\left(f(z)\partial_u+i\beta^a\partial_a f(z)-\frac{i\kappa}{4}\partial_a \partial_a f(z)\right)O_X.
	\end{equation}
	A local function $f(w)$ acts through the nonlocal Poisson transform $\mathcal{I}_{A}^{g}[f]$. \par

	At subleading order, the soft factor contains the orbital Lorentz generator,
	\begin{equation}
		S_{g,A}^{(1),h}(w)=i\frac{p_A\cdot\varepsilon_h(w)}{p_A\cdot q_0(w)}q_{0\rho}(w)\varepsilon_{h\nu}(w)J_A^{\rho\nu}.
	\end{equation}
	In a soft matrix element, the $\Delta=0$ residue gives the corresponding Carrollian current together with a contribution from the leading pole,
	\begin{equation}
		J_h^{(1)}(w)=\lim_{\omega\to0}(1+\omega\partial_\omega)a_h^{\mathrm{out}}(\omega,w)=\lim_{\Delta\to0}\Delta\mathcal{O}_{\Delta,h}(u,w)+2iuJ_h(u,w).
	\end{equation}
	For a superrotation parameter $Y^a(w)$, the subleading graviton smearing kernel is
	\begin{equation}
		K_Y^h(w)=-\frac{1}{32\pi}(\partial_1-i\,\operatorname{sgn}(h)\partial_2)^3(Y^1(w)+i\,\operatorname{sgn}(h)Y^2(w)).
	\end{equation}
	Using $J_h^{(1)}$ in place of $J_h$ and $K^h=K_Y^h$ in eq. \eqref{eq:SoftCharge} gives the superrotation soft charge $Q_S^{(1)}[Y]$.
	For a local superrotation parameter $Y^a(w)$, the subleading soft theorem gives the following action
	\begin{equation}
		\mathcal{I}_{A}^{(1)}[Y]=\eta_A\sum_{h=\pm2}\int \dd[2]{w}\,K_Y^h(w)S_{g,A}^{(1),h}(w).
	\end{equation}
	The global vectors listed below are at most quadratic in the celestial coordinates, so $K_Y^h$ vanishes for these modes.
	For these global modes, the hard action instead satisfies $\mathcal{I}_{A}^{(1)}[Y]O_X=i\eta_A\delta_YO_X$, where
	\begin{equation}
		\begin{aligned}
			\delta_YO_X
			&=
			\Big(
			Y^a\partial_{z^a}
			+\lambda(u\partial_u+\kappa\partial_\kappa+\Delta) +
			(\partial_b Y_a-\lambda\delta_{ab})\beta^b\partial_{\beta^a} +
			\frac{\kappa}{2}\partial_a \lambda\,\partial_{\beta^a}
			+iu\beta^a\partial_a \lambda
			\Big)O_X,
		\end{aligned}
	\end{equation}
	Here $\lambda=\frac{1}{2}\partial_a Y^a$, and the spin term is absent because the hard legs are scalar.
	The six global vectors are
	\begin{equation}
		\begin{aligned}
			&Y_{P}^{a}=c^a,\\
			&Y_{D}^{a}=z^a, \qquad Y_{J}^{a}=(-z^2,z^1),\\
			&Y_{K}^{a}=2(b\cdot z)z^a-b^a\vec{z}^{\,2}.
		\end{aligned}
	\end{equation}
	Together, these six generators realize the bulk Lorentz rotations as boundary Carrollian conformal actions.
	The soft charge $Q_S^{(1)}[Y]$ therefore vanishes for these global vectors.
	The Ward identity then contains only the hard charge and reduces to Lorentz invariance of the amplitude.
	\par

	The distinction between global and local transformations follows from the null frames used in the dictionary.
	The soft charge inserts a massless radiative soft mode for both massless and massive hard legs, so the difference arises from the hard action.
	For a massless hard leg, the angular pole in the soft factor localizes the smearing at its celestial position.
	For a massive hard leg, the momentum has nonzero projections onto all null frames, and the hard action remains an integral over the entire celestial sphere.
	The global modes combine these projections to reproduce the ordinary $U(1)$ and Poincar\'e actions.
	The hard actions generated by local $\alpha(w)$ and $f(w)$ retain nonlocal angular integrals, and the frame picture indicates an analogous structure for local superrotations. \par

\section{Discussion}

	Our construction shows how massive one-particle states in a 4D bulk can be described on its 3D null boundary without assigning their timelike worldlines an endpoint on $\mathscr I^\pm$.
	The $\kappa\neq0$ orbits of the complete Carrollian representation carry the nonzero quadratic Poincar\'e Casimir required by a positive mass.
	Each celestial point specifies a null frame, and the collection of momentum projections over the celestial sphere encodes the timelike momentum.
	Together with the standard radiative dictionary for massless particles, our construction allows massive and massless external states to be described within a common Carrollian framework on $\mathscr I^\pm$. \par

	The normalized dictionary gives an isometric embedding of the bulk Hilbert space into the boundary state space, and the reconstruction map is its Hermitian adjoint.
	On the unitary line $\Delta=2+i\nu$ with $\nu\in\mathbb{R}$, the two maps satisfy $\hat{\mathcal{G}}^\dagger\hat{\mathcal{G}}=\mathbf{I}_{\mathcal{H}_{j}^{(+)}}$ on $\mathcal{H}_{j}^{(+)}$ and $\hat{\mathcal{G}}\hat{\mathcal{G}}^\dagger=\Pi_{\mathcal{S}_O^{>0}}$ on $\mathcal{S}_O^{>0}$.
	Through the dictionary, the complete local Carrollian operators $O_X$ form an overcomplete basis for $\mathcal{H}_{j}^{(+)}$.
	In a holographic realization, the projection operator $\Pi$ selects the combinations that encode physical massive bulk states.
	For arbitrary spin, the scalar support and the Hermitian adjoint relation between the two maps remain unchanged, while a unitary spin matrix factor carries the spin dependence, as shown in Appendix \ref{app:GeneralSpin}. \par

	For the scalar example considered here, applying the dictionary to the K\"all\'en--Lehmann representation shows that it carries bulk dynamics to the boundary.
	The bulk spectral density fixes the coefficient function of the generic non-contact Carrollian two-point function.
	For correlators arising from the bulk, this relation restricts the functional freedom allowed by Carrollian symmetry to data with a spectral interpretation.
	This shows how physical consistency conditions in the bulk constrain Carrollian data beyond the Ward identities. \par

	The analysis based on soft theorems gives a second physical consequence for scalar massive hard legs.
	For a massive hard leg, the timelike momentum has nonzero projections at every celestial point, so the actions of local large gauge transformations and supertranslations remain angular integrals, while the same projection picture indicates an analogous structure for local superrotations.
	The global modes instead combine the angular dependence to reproduce the ordinary $U(1)$ and Poincar\'e actions. \par

	The null frame construction also admits a counterpart for a 3D bulk and a 2D boundary.
	Appendix \ref{app:3DMassiveDictionary} gives the corresponding relation between bulk and boundary symmetries, together with the normalized dictionary and reconstruction maps. \par

	Several questions follow from this construction.
	First, it remains to determine whether and in what sense the massive dictionary can be related to the standard massless collinear dictionary \cite{Nguyen:2023miw,Furugori:2023hgv}, and whether an $m\to0$ limit connects the two after normalization is treated consistently.
	A second question is how the present null frame dictionary is related to the massive celestial conformal primary basis and to Carrollian descriptions at the blow-up of timelike infinity.
	It remains to determine whether these constructions provide different boundary realizations of the same massive Hilbert space \cite{Pasterski:2017kqt,Donnay:2022wvx,Iacobacci:2024laa,Have:2024dff,Law:2020tsg,Narayanan:2020amh,Hao:2023wln,Banerjee:2024yir,Liu:2026toc}.
	Finally, broader questions concern what further constraints can be imposed on the boundary theory, which boundary theory provides the holographic dual of an asymptotically flat bulk, and how its operator content and dynamics can be characterized \cite{Donnay:2022aba,Alday:2024yyj,Fredenhagen:2026pia,Ruzziconi:2026bix,deBoer:2003vf,Marolf:2006bk,Hijano:2019qmi,Laddha:2022nmj,Jain:2023fxc,H:2024cfo,Navarro:2025xln,Melton:2026dbf,Navarro:2026rna}. \par

\section*{Acknowledgments}

	We thank Bin Chen, Reiko Liu, Jiang Long, and Wen-Jie Ma for valuable discussions. \par

\appendix

\section{Bulk and Boundary Coordinates and Symmetries}\label{app:BulkBoundaryParameters}

	To relate the bulk and boundary coordinates, we parameterize a point near future null infinity by $(t,r,\theta,\phi)$ as
	\begin{equation}
		x^\mu=(t,r\sin{\phi}\cos{\theta},r\sin{\phi}\sin{\theta},r\cos{\phi}).
	\end{equation}
	The Carrollian coordinates used in the letter are
	\begin{equation}
		u=\frac{t-r}{1+\cos{\phi}},
		\qquad
		z^1=\cos{\theta}\tan{\frac{\phi}{2}},
		\qquad
		z^2=\sin{\theta}\tan{\frac{\phi}{2}}.
	\end{equation}
	Here $\phi$ is the polar angle and $\theta$ is the azimuthal angle.
	Future null infinity is obtained by taking $r\to\infty$ at fixed $(u,z^a)$.
	The construction at past null infinity follows analogously using advanced coordinates and the antipodal identification. \par

	The null frame associated with each celestial point $(z^1,z^2)$ is given in eq. \eqref{eq:NullFrame}.
	The vectors $q_0$ and $q_3$ form a null vector pair, while $q_1$ and $q_2$ are transverse spacelike vectors.
	Their inner products are
	\begin{equation}
		\begin{aligned}
			& q_0^2=q_3^2=0,
			&& q_0\cdot q_3=-1,
			&&q_a\cdot q_b=\delta_{ab},
			&& q_0\cdot q_a=q_3\cdot q_a=0.
		\end{aligned}
	\end{equation}
	The corresponding completeness relation is
	\begin{equation}
		\eta^{\mu\nu}=-q_0^\mu q_3^\nu-q_3^\mu q_0^\nu+\delta^{ab}q_a^\mu q_b^\nu.
	\end{equation}
	Together with the delta function support of the scalar kernel, this frame gives the momentum reconstruction in eq. \eqref{eq:MomentumFrame}. \par

	For the bulk generators, we use the vector field convention
	\begin{equation}
		\begin{aligned}
			\xi_{\mathbb{P}^0}&=-\partial_0,
			&\xi_{\mathbb{P}^i}&=\partial_i,\\
			\xi_{\mathbb{J}^{0i}}&=x^0\partial_i+x^i\partial_0,
			&\xi_{\mathbb{J}^{ij}}&=x^i\partial_j-x^j\partial_i.
		\end{aligned}
	\end{equation}
	Here $\mathbb{P}^\mu$ and $\mathbb{J}^{\mu\nu}$ denote the bulk Poincar\'e generators.
	Pushing these vector fields to null infinity and using $\partial_a=\partial/\partial z^a$ gives the global Carrollian generators
	\begin{equation}
		\begin{aligned}
			\xi_{P^0}&=\partial_u,
			&\xi_{P^a}&=\partial_a,\\
			\xi_D\,&=u\partial_u+z^a\partial_a,
			&\xi_{J^{12}}&=z^1\partial_2-z^2\partial_1,
			\qquad\xi_{B^a}=z^a\partial_u,\\
			\xi_{K^0}&=-\vec{z}^{\,2}\partial_u,
			&\xi_{K^a}&=2uz^a\partial_u+(2z^az^b-\vec{z}^{\,2}\delta^{ab})\partial_b.
		\end{aligned}
	\end{equation}
	The resulting one-to-one correspondence between the bulk Poincar\'e generators and the boundary Carrollian generators is
	\begin{equation}
		\begin{aligned}
			\mathbb{J}^{03}&=-D,
			&\mathbb{J}^{12}&=J^{12},
			&\mathbb{J}^{0a}&=\frac{1}{2}(P^a-K^a),
			&\mathbb{J}^{a3}&=-\frac{1}{2}(P^a+K^a),\\
			\mathbb{P}^0&=\frac{1}{2}(K^0-P^0),
			&\mathbb{P}^a&=-B^a,
			&\mathbb{P}^3&=-\frac{1}{2}(K^0+P^0).
		\end{aligned}
	\end{equation}
	This invertible change of basis shows explicitly that the global Carrollian action on $(u,z^a)$ is the boundary limit of the bulk Poincar\'e action. \par

\section{Spin Matrix Factors}\label{app:GeneralSpin}

	Let $\mathcal{G}(p;X)$ and $\mathcal{K}(X;p)$ denote the scalar kernels in Section \ref{sec:MassiveDictionary}.
	For spin $j$, the rows of the spin matrix are labeled by $l=j,j-1,\ldots,-j$, while its columns are labeled by $s=j,j-1,\ldots,-j$.
	The dictionary and reconstruction kernels take the form
	\begin{equation}
		\begin{aligned}
			\mathcal{G}^{(j)}_{ls}(p;X)
			={}&2m\,e^{2iu\,p\cdot q_3(z)}\abs{\kappa}^{1-\Delta}
			\mathcal{R}^{(j)}_{ls}
			\delta^{(2)}\!\left(\beta^a+p\cdot q_a(z)\right)
			\delta\!\left(\kappa+p\cdot q_0(z)\right),\\
			\mathcal{K}^{(j)}_{sl}(X;p)
			={}&2m\,e^{-2iu\,p\cdot q_3(z)}\abs{\kappa}^{\Delta-3}
			(\mathcal{R}^{(j)\dagger})_{sl}
			\delta^{(2)}\!\left(\beta^a+p\cdot q_a(z)\right)
			\delta\!\left(\kappa+p\cdot q_0(z)\right).
		\end{aligned}
	\end{equation}
	Thus the scalar support is unchanged, and all spin dependence is carried by $\mathcal{R}^{(j)}$. \par

	For compactness, define $p_\pm=p^1\pm ip^2$, $\beta_\pm=\beta^1\pm i\beta^2$, and $\chi_p=p^0+p^3+m$.
	For spin $\frac{1}{2}$, the matrices are
	\begin{equation}
		\begin{aligned}
			\mathsf{F}^{(\frac{1}{2})}
			&=
			\begin{pmatrix}
				1 & \beta_-/m\\
				-\beta_+/m & 1
			\end{pmatrix},\qquad
			\mathsf{M}^{(\frac{1}{2})}
			&=
			\begin{pmatrix}
				p_- & -\chi_p\\
				\chi_p & p_+
			\end{pmatrix}.
		\end{aligned}
	\end{equation}
	The normalization coefficient is
	\begin{equation}
		c_{\frac{1}{2}}=\frac{m}{\sqrt{2(p^0+p^3)(p^0+m)(m^2+\vec{\beta}^{\,2})}}.
	\end{equation}
	The spin matrix factor is then
	\begin{equation}
		\mathcal{R}^{(\frac{1}{2})}=c_{\frac{1}{2}}\mathsf{F}^{(\frac{1}{2})}\mathsf{M}^{(\frac{1}{2})}.
	\end{equation}
	For spin $1$, the corresponding matrices are
	\begin{equation}
		\mathsf{F}^{(1)}=
		\begin{pmatrix}
			1 & \sqrt{2}\,\beta_-/m & \beta_-^2/m^2\\
			-\sqrt{2}\,\beta_+/m & (m^2-\vec{\beta}^{\,2})/m^2 & \sqrt{2}\,\beta_-/m\\
			\beta_+^2/m^2 & -\sqrt{2}\,\beta_+/m & 1
		\end{pmatrix},
	\end{equation}
	\begin{equation}
		\mathsf{M}^{(1)}=
		\begin{pmatrix}
			p_-^2 & -\sqrt{2}\,p_-\chi_p & \chi_p^2\\
			\sqrt{2}\,p_-\chi_p & p_+p_--\chi_p^2 & -\sqrt{2}\,p_+\chi_p\\
			\chi_p^2 & \sqrt{2}\,p_+\chi_p & p_+^2
		\end{pmatrix}.
	\end{equation}
	The normalization coefficient is
	\begin{equation}
		c_1=\frac{m^2}{2(p^0+p^3)(p^0+m)(m^2+\vec{\beta}^{\,2})}.
	\end{equation}
	The spin matrix factor is then
	\begin{equation}
		\mathcal{R}^{(1)}=c_1\mathsf{F}^{(1)}\mathsf{M}^{(1)}.
	\end{equation}
	A direct calculation gives $\mathcal{R}^{(j)}\mathcal{R}^{(j)\dagger}=\mathbf{1}_{2j+1}$ for $j=\frac{1}{2},1$. \par

	For generic spin $j$, the matrices follow directly from the spin-$j$ polynomial realization of $\mathrm{SU}(2)$.
	For $l,s=j,j-1,\ldots,-j$, their matrix elements are
	\begin{equation}
		\begin{aligned}
			\mathsf{F}^{(j)}_{ls}
			={}&\sqrt{\frac{\binom{2j}{j+s}}{\binom{2j}{j+l}}}
			\sum_{k=\max(0,l+s)}^{\min(j+s,j+l)}
			(-1)^{j+s-k}
			\binom{j+s}{k}\binom{j-s}{j+l-k}
			\times
			\left(\frac{\beta_+}{m}\right)^{j+s-k}
			\left(\frac{\beta_-}{m}\right)^{j+l-k}.
		\end{aligned}
	\end{equation}
	\begin{equation}
		\begin{aligned}
			\mathsf{M}^{(j)}_{ls}
			={}&\sqrt{\frac{\binom{2j}{j+s}}{\binom{2j}{j+l}}}
			\sum_{k=\max(0,l+s)}^{\min(j+s,j+l)}
			(-1)^{j+l-k}
			\binom{j+s}{k}\binom{j-s}{j+l-k}\times
			p_-^{k}p_+^{k-l-s}\chi_p^{2j+l+s-2k}.
		\end{aligned}
	\end{equation}
	In both expressions, the sums run over integers, and every exponent is a nonnegative integer over the stated ranges.
	The normalized product gives each element of the spin matrix factor as
	\begin{equation}
		\begin{aligned}
			&\mathcal{R}^{(j)}_{ls}=c_j\sum_{r=-j}^{j}\mathsf{F}^{(j)}_{lr}\mathsf{M}^{(j)}_{rs},\\
			&c_j=\left(\frac{m^2}{2(p^0+p^3)(p^0+m)(m^2+\vec{\beta}^{\,2})}\right)^j.
		\end{aligned}
	\end{equation}
	The matrices satisfy
	\begin{equation}
		\begin{aligned}
			\mathsf{F}^{(j)}\mathsf{F}^{(j)\dagger}
			&=\left(\frac{m^2+\vec{\beta}^{\,2}}{m^2}\right)^{2j}\mathbf{1}_{2j+1},\\
			\mathsf{M}^{(j)}\mathsf{M}^{(j)\dagger}
			&=(2(p^0+p^3)(p^0+m))^{2j}\mathbf{1}_{2j+1},\\
			\mathcal{R}^{(j)}\mathcal{R}^{(j)\dagger}
			&=\mathbf{1}_{2j+1}.
		\end{aligned}
	\end{equation}
	Thus the spin matrix factor is unitary and regular on the positive energy branch of the massive mass shell. \par

	On the unitary line $\Delta=2+i\nu$, the exponential factors and the powers of $\abs{\kappa}$ in the two kernels are complex conjugates, while the delta functions are real.
	For spin $j$, the explicit Hermitian conjugation of the spin matrix factor therefore gives
	\begin{equation}
		\mathcal{K}^{(j)}_{sl}(X;p)
		=(\mathcal{G}^{(j)}_{ls}(p;X))^*,
		\qquad
		\hat{\mathcal{K}}=\hat{\mathcal{G}}^\dagger,
		\qquad
		j\in\frac{1}{2}\mathbb{Z}_{\geq0}.
	\end{equation}
	Thus, for arbitrary spin, the reconstruction kernel is obtained from the dictionary kernel by complex conjugation and exchange of the spin indices. \par

\section{Massive Dictionary in 3D}\label{app:3DMassiveDictionary}

	This appendix gives the counterpart of the construction in Section \ref{sec:MassiveDictionary} for a 3D bulk and a 2D boundary \cite{Dutta:2026etj,Hao:2025btl,Hao:2026cqm}.
	We use the metric $\eta_{\mu\nu}=\mathrm{diag}(-,+,+)$ and parameterize the bulk coordinates by
	\begin{equation}
		x^\mu=(t,r\cos{\theta},r\sin{\theta}).
	\end{equation}
	The Carrollian coordinates at future null infinity are
	\begin{equation}
		u=\frac{t-r}{1+\cos{\theta}},
		\qquad
		z=\tan{\frac{\theta}{2}},
	\end{equation}
	with $r\to\infty$ taken at fixed $(u,z)$.
	The 3D null frame associated with the celestial coordinate $z$ is given by
	\begin{equation}
		\begin{aligned}
			q_0^\mu(z)&=(1+z^2,1-z^2,2z),\\
			q_1^\mu(z)&=\frac{1}{2}\partial_z q_0^\mu(z)=(z,-z,1),\\
			q_2^\mu(z)&=\frac{1}{4}\partial_z^2 q_0^\mu(z)=\frac{1}{2}(1,-1,0).
		\end{aligned}
	\end{equation}
	Here $q_0$ and $q_2$ are null, while $q_1$ is spacelike, and their nonzero inner products are $q_0\cdot q_2=-1$ and $q_1^2=1$.
	They obey the completeness relation
	\begin{equation}
		\eta^{\mu\nu}=-q_0^\mu q_2^\nu-q_2^\mu q_0^\nu+q_1^\mu q_1^\nu.
	\end{equation}
	Thus the same null frame mechanism used in the letter applies in 3D, where each celestial point is labeled by $z$. \par

	For completeness, we record the relation between bulk and boundary symmetry generators.
	The 3D Poincar\'e vector fields are
	\begin{equation}
		\begin{aligned}
			\xi_{\mathbb{P}^0}&=-\partial_0,
			&\xi_{\mathbb{P}^1}&=\partial_1,
			&\xi_{\mathbb{P}^2}&=\partial_2,\\
			\xi_{\mathbb{J}^{01}}&=x^0\partial_1+x^1\partial_0,
			&\xi_{\mathbb{J}^{02}}&=x^0\partial_2+x^2\partial_0,
			&\xi_{\mathbb{J}^{12}}&=x^1\partial_2-x^2\partial_1.
		\end{aligned}
	\end{equation}
	Their boundary limits are generated by
	\begin{equation}
		\begin{aligned}
			\xi_{P^0}&=\partial_u,
			&\xi_{P^1}&=\partial_z,
			&\xi_{B^1}&=z\partial_u,\\
			\xi_D&=u\partial_u+z\partial_z,
			&\xi_{K^0}&=-z^2\partial_u,
			&\xi_{K^1}&=2uz\partial_u+z^2\partial_z.
		\end{aligned}
	\end{equation}
	The bulk and boundary bases are related by
	\begin{equation}
		\begin{aligned}
			\mathbb{J}^{01}&=-D,
			&\mathbb{J}^{02}&=\frac{1}{2}(P^1-K^1),
			&\mathbb{J}^{12}&=\frac{1}{2}(P^1+K^1),\\
			\mathbb{P}^0&=\frac{1}{2}(K^0-P^0),
			&\mathbb{P}^1&=-\frac{1}{2}(K^0+P^0),
			&\mathbb{P}^2&=-B^1.
		\end{aligned}
	\end{equation}
	This change of basis identifies the 3D Poincar\'e symmetry generators with the global Carrollian conformal generators. \par

	At fixed spin $j$ for a massive particle, let $\mathcal{H}_{j,\mathrm{3D}}^{(+)}$ denote the outgoing bulk space and let $X=(u,z;\Delta,\beta,\kappa)$ label the complete boundary representation.
	The unitary line is $\Delta=\frac{3}{2}+i\nu$ with $\nu\in\mathbb{R}$.
	The $\kappa>0$ and $\kappa<0$ orbits describe outgoing and incoming states, respectively, and the former spans $\mathcal{S}_{O,\mathrm{2D}}^{>0}$. \par

	Using our momentum space convention, the outgoing dictionary is
	\begin{equation}
		\ket{O_X}=\hat{\mathcal{G}}\ket{p;j}=\int_{\substack{p^2<0\\p^0>0}}\frac{\dd[3]{p}}{(2\pi)^{3/2}}\,\ket{p;j}\mathcal{G}^{(j)}(p;X),
	\end{equation}
	where $m^2=-p^2$ and
	\begin{equation}
		\begin{aligned}
			\mathcal{G}^{(j)}(p;X)
			={}&\left(\frac{8m^2}{\pi}\right)^{1/4}
			e^{2iu\,p\cdot q_2}
			\abs{\kappa}^{1-\Delta}
			\mathcal{R}^{(j)}
			\times\delta\!\left(\beta+p\cdot q_1(z)\right)
			\delta\!\left(\kappa+p\cdot q_0(z)\right).
		\end{aligned}
	\end{equation}
	Because the $\mathrm{SO}(2)$ little group representation is one dimensional, the spin matrix factor reduces to the phase
	\begin{equation}
		\mathcal{R}^{(j)}=
		\left(
			\frac{\kappa+(m+i\beta)(1+iz)}{\kappa+(m-i\beta)(1-iz)}
		\right)^j,
		\qquad
		\abs{\mathcal{R}^{(j)}}=1.
	\end{equation}
	The two support conditions give
	\begin{equation}
		\beta=-p\cdot q_1(z),
		\qquad
		\kappa=-p\cdot q_0(z),
	\end{equation}
	and therefore reconstruct the momentum as
	\begin{equation}
		p^\mu=\frac{m^2+\beta^2}{2\kappa}q_0^\mu(z)-\beta q_1^\mu(z)+\kappa q_2^\mu(z).
	\end{equation}
	Thus $\beta$ and $\kappa$ again record the projections of a timelike momentum onto the null frame without assigning a boundary endpoint to the massive trajectory. \par

	The reconstruction map is
	\begin{equation}
		\ket{p;j}=\hat{\mathcal{K}}\ket{O_X}=\int_{\kappa>0}\dd{u}\dd{z}\dd{\beta}\dd{\kappa}\,\ket{O_X}\mathcal{K}^{(j)}(X;p),
	\end{equation}
	with reconstruction kernel
	\begin{equation}
		\begin{aligned}
			\mathcal{K}^{(j)}(X;p)
			={}&\left(\frac{8m^2}{\pi}\right)^{1/4}
			e^{-2iu\,p\cdot q_2}
			\abs{\kappa}^{\Delta-2}
			(\mathcal{R}^{(j)})^{-1}
			\times\delta\!\left(\beta+p\cdot q_1(z)\right)
			\delta\!\left(\kappa+p\cdot q_0(z)\right).
		\end{aligned}
	\end{equation}
	We normalize the momentum eigenstates by $\bra{p^\prime;j^\prime}\ket{p;j}=(2\pi)^{3/2}\delta^{(3)}(p-p^\prime)\delta_{j^\prime j}$.
	Since $\mathcal{R}^{(j)}$ is a phase and $\Delta=\frac{3}{2}+i\nu$ with $\nu\in\mathbb{R}$ on the unitary line, the two kernels satisfy
	\begin{equation}
		\mathcal{K}^{(j)}(X;p)=\mathcal{G}^{(j)}(p;X)^*,
		\qquad
		\hat{\mathcal{K}}=\hat{\mathcal{G}}^\dagger.
	\end{equation}
	Contracting the dictionary with the bulk bra state and the reconstruction with the boundary bra state gives the overlap relations
	\begin{equation}
		\mathcal{G}^{(j)}(p;X)=\bra{p;j}\ket{O_X},
		\qquad
		\mathcal{K}^{(j)}(X;p)=\bra{O_X}\ket{p;j}.
	\end{equation}
	They are therefore state overlaps rather than bulk-to-boundary propagators, since massive worldlines do not reach null infinity.
	Their integral form is nevertheless analogous to an HKLL map.
	The chosen normalization gives the completeness identity
	\begin{equation}
		(2\pi)^{3/2}\delta^{(3)}(p-p^\prime)=\int_{\kappa>0}\dd{u}\dd{z}\dd{\beta}\dd{\kappa}\,
		\mathcal{K}^{(j)}(X;p^\prime)\mathcal{G}^{(j)}(p;X).
	\end{equation}
	Consequently, the two compositions satisfy
	\begin{equation}
		\hat{\mathcal{G}}^\dagger\hat{\mathcal{G}}=\mathbf{I}_{\mathcal{H}_{j,\mathrm{3D}}^{(+)}},
		\qquad
		\hat{\mathcal{G}}\hat{\mathcal{G}}^\dagger=\Pi_{\mathcal{S}_{O,\mathrm{2D}}^{>0}}\neq\mathbf{I}_{\mathcal{S}_{O,\mathrm{2D}}^{>0}}.
	\end{equation}
	For each bulk momentum, the continuous choice of null frame supplies redundant boundary labels, so the boundary states are overcomplete. \par

\bibliographystyle{JHEP}
\bibliography{refs}

\end{document}